\documentclass{roadef}

\usepackage{wrapfig}
\usepackage{graphicx}

\begin{document}

% Le titre du papier
\title{Planification d'observations sur télescope avec nuits incertaines}

% Le titre court
\def\shorttitle{$P_m|r_j|\sum w_j U_j$ avec $m$ incertain}

% Les auteurs et leur numéro d'affiliation
\author{Thomas Rahab Lacroix, Pierre Lemaire}

% Les affiliations (par ordre croissant des numéros d'affiliation) séparées par \and
\institute{
Univ. Grenoble Alpes, CNRS, Grenoble INP, G-SCOP, 38000 Grenoble, France\\
\email{\{thomas.rahab-lacroix, pierre.lemaire\}@grenoble-inp.fr}
}

% Création de la page de titre
\maketitle
\thispagestyle{empty}

% Les mots-clés
\keywords{Astronomie, optimisation stochastique, horizon glissant, exoplanètes}

\section{Le problème des astronomes}

Nous proposons un problème d'ordonnancement original issu d'une collaboration avec des astronomes de l'Institut de Planétologie et d'Astronomie de Grenoble. Pour faire de l'imagerie directe d'exoplanètes, les télescopes sont rares ; ainsi le VLT (Very Large Telescope) est un des seuls télescopes européens adapté. Cette ressource rare et les très nombreuses cibles célestes intéressantes obligent à calculer un ordonnancement optimisé.

Concrètement, on dispose de $m$ nuits pour observer un ensemble de cibles. Chaque cible $j$ est qualifiée par un intérêt à priori ($w_j$), une durée d'observation ($processing$ $time$ $p_j$) ainsi qu'une fenêtre de visibilité ($release$ $date$ $r_j$, $due$ $date$ $d_j$). Toutes les nuits sont identiques et il faut maximiser l'intérêt global, le gain. On a donc un problème $P_m|r_j|\sum w_j U_j$. 

Ce problème d'ordonnancement déterministe, fortement NP-complet, a déjà été bien étudié (par exemple \cite{maxi_throughtput} encore récemment) tandis qu'une version spécifique aux astronomes avec des nuits distinctes a été étudiée dans \cite{star_schedule}.

Ici, nous considérons un élément nouveau : la météo. La formulation choisie est une simplification de la réalité où l’on considère qu’une nuit est soit parfaite (toutes les observations sont faisables), soit le télescope est fermé. On a alors un problème $P_m |r_j| \sum w_j U_j$ avec $m$ incertain. C'est seulement au début de chaque nuit que l'on sait si elle est parfaite ou non. 

%Une remarque intéressante est que pour un ordonnancement donné, l'ordre dans lequel réaliser ses nuits est évident. Effectivement, dès qu'une nuit est observable, planifier à chaque fois la nuit de l'ordonnancement rapportant le plus, domine tout autre arrangement.

Nous proposons et discutons différentes approches qui correspondent à autant de visions différentes pour gérer cette incertitude.

\section{Approches de planifications avec nuits incertaines}

\begin{figure}[h]
    \begin{minipage}{0.5\textwidth}
        \centering
        \includegraphics[width=\linewidth]{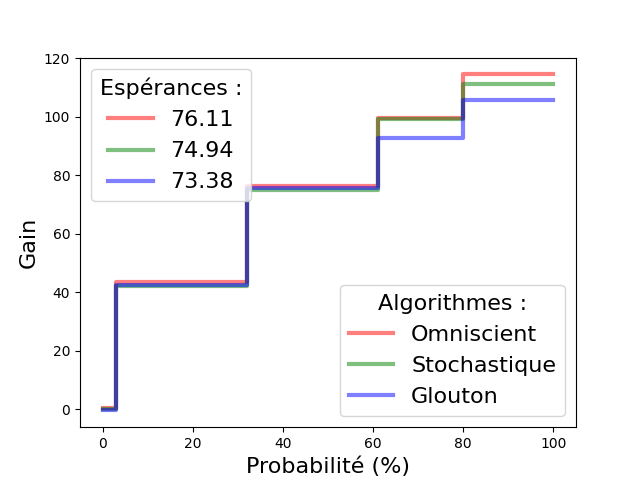}
        \caption{Comparaison des trois approches}
        \label{fig:compa_3_algos}
    \end{minipage}%
    \hfill
    \begin{minipage}{0.5\textwidth}
        \centering
        \includegraphics[width=\linewidth]{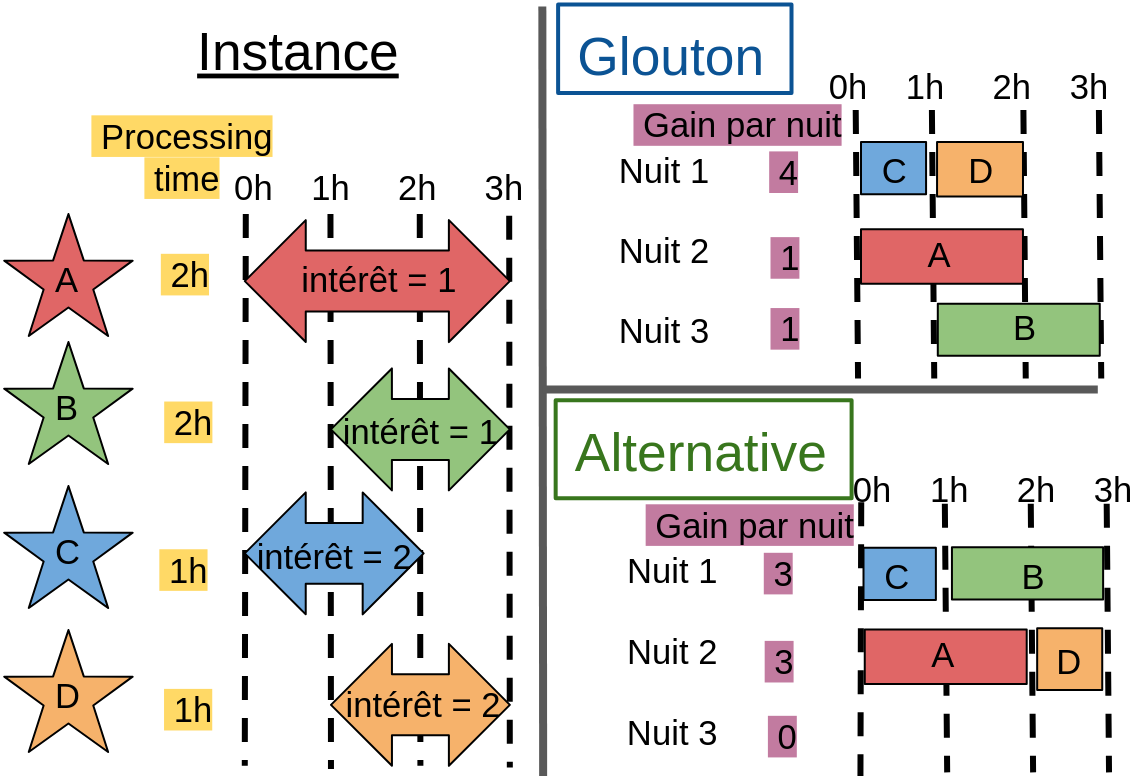}
        \caption{Limite de Glouton}
        \label{fig:greedy_not_opti}
    \end{minipage}
\end{figure}

Pour toutes les approches, l'ordonnancement est calculé sur les $m$ nuits potentielles. Si seulement $k$ nuits sont parfaites, alors la fin de la planification n'est pas utilisée. La Figure~\ref{fig:compa_3_algos} illustre une instance théorique sur quatre nuits.
Chaque marche correspond à une nuit de plus d'observation et sa longueur est la probabilité que cela arrive (obtenue via des prévisions météo). Ainsi, comme le gain est en ordonnées, l’aire sous une courbe est l'espérance, le gain réalisé en moyenne. %Les courbes de la Figure~\ref{fig:compa_3_algos} peuvent aussi être vues comme les réciproques des fonctions de répartitions du gain.

Approche 1 : Glouton. Cela ressemble à ce que les astronomes font actuellement. On planifie chaque nuit comme si c'était la dernière, avec ce qu'il y a de mieux dans ce qui reste à faire. Cette heuristique peut facilement se faire piéger, comme l'illustre la Figure~\ref{fig:greedy_not_opti} où Glouton met trois nuits à planifier toutes les observations alors que la solution alternative n'en a besoin que de deux. Ainsi si la probabilité que deux nuits soient observables est très élevée, la Figure~\ref{fig:greedy_not_opti} montre que Glouton a des limites et l'améliorer est bénéfique pour les astronomes. 

Approche 2 : Stochastique. Via un solveur (MILP ou CP), on cherche la solution qui maximise l’espérance. Sur la Figure~\ref{fig:compa_3_algos}, on observe la courbe Stochastique majoritairement au dessus de Glouton ainsi que son espérance supérieure. Cela prouve l'intérêt de cette approche.

Approche 3 : Omniscient. Contrairement aux deux autres, sa courbe sur la Figure~\ref{fig:compa_3_algos} ne représente pas une solution. Omniscient n’a plus de notion d’incertitude et chaque marche est la solution optimale pour chaque nombre de nuits possibles. L’écart entre la courbe Omniscient et Stochastique illustre donc le coût de l’incertitude.

\section{Version réactive du problème}

%Dans les faits, les nuits sont séparées de 24h et des données incertaines sont dévoilées. On utilise alors une méthode d’horizon glissant [1].
%On sait après chaque nuit si elle a été parfaite ou pas, ce qui permet de mettre à jour la planification pour les nuits restantes selon une méthode d'horizon glissant \cite{rolling_hori}.
Il semble pertinent de s'adapter entre les nuits en fonction de leurs observabilités via une méthode d'horizon glissant \cite{rolling_hori}.
A chaque nuit se révélant observable, les probabilités sont mises à jour, l'optimum stochastique est recalculé et la première nuit du nouveau planning est exécutée et ce ainsi de suite. On calcule alors l'espérance réactive de l'ensemble de cette simulation. Cette espérance est toujours supérieure à celle de Stochastique, preuve de l'intérêt d'une approche réactive. La Table~\ref{tab:espe_react_compa} montre ce que fait gagner la méthode Réactive par rapport à la méthode Stochastique pour des instances de quatre nuits. Actuellement de l'ordre du pourcent, cette amélioration pourrait devenir bien plus conséquente sur de grandes instances.

\begin{table}[h!]
    \centering
    \begin{tabular}{c|c|c|c}
    Instance & Espérance Réactive & Espérance Stochastique & Amélioration \\\hline
    1 & 70.81 & 70.56 & 0.35\% \\
    2 & 74.60 & 74.35 & 0.33\% \\
    3 & 78.36 & 78.11 & 0.32\%
    \end{tabular}
    \caption{\label{tab:espe_react_compa}Comparaison des espérances Réactive et Stochastique}
\end{table}

\section{Conclusions et perspectives}

Cette étude présente un problème neuf et des techniques pour l'aborder. On découvre l'intérêt d'utiliser des méthodes élaborées, une mesure de l'incertitude et l'amélioration par le réactif. Les perspectives sont de complexifier le modèle pour se rapprocher du problème réel des astronomes.

% La bibliographie

\bibliographystyle{plain}

% Version "on-line" de la bibliographie, mais il est
% également possible d'utiliser un fichier ".bib" et d'utiliser BibTeX

\end{document}